\documentclass{birkmult}
 \newtheorem{thm}{Theorem}[section]
 
 \newtheorem{lem}[thm]{Lemma}
 \newtheorem{prop}[thm]{Proposition}
 \theoremstyle{definition}
 \newtheorem{defn}[thm]{Definition}
 \theoremstyle{remark}
 \newtheorem{rem}[thm]{Remark}
 
 \numberwithin{equation}{section}

\def\sqr#1#2{{\vcenter{\vbox{\hrule height.#2pt\hbox{\vrule width.#2pt
height#1pt \kern#1pt \vrule width.#2pt}\hrule height.#2pt}}}}
\def\d{\partial}

\def\=d{\,{\buildrel\rm def\over =}\,}

\def\1{{\leavevmode{\rm 1\ifmmode\mkern -4.8mu\else\kern -.3em\fi l}}}
\newcommand{\del}{\partial}

\newcommand{\Rcl}{R_\mathrm{cl}}
\newcommand{\ets}{e_\otimes^S}

\newcommand{\defi}{\stackrel{\scriptscriptstyle{\mathrm{def}}}{=}}

\newcommand{\dA}{\delta_A}
\newcommand{\ffgx}{\frac{\delta S_0}{\delta \varphi(x)}}

\renewcommand{\theequation}{\thesection.\arabic{equation}}
\newcommand{\beq}{\begin{equation}}
\newcommand{\eeq}{\end{equation}}
\newcommand{\bg}{\begin{gather}}
\newcommand{\eg}{\end{gather}}
\newcommand{\CC}{\mathbb C}
\newcommand{\RR}{\mathbb R}

\newcommand{\NN}{\mathbb N}
\newcommand{\MM}{\mathbb M}
\newcommand{\TT}{\mathbb T}
\newcommand{\DD}{\mathbb D}
\newcommand{\supp}{{\rm supp\>}}

\renewcommand{\labelenumi}{(\roman{enumi})}

\begin{document}
%
%
%
%
%
%
%
%
%
\title[Quantum Action Principle in Causal Perturbation Theory]
 {The Quantum Action Principle in the \\ framework of Causal Perturbation Theory}
\author[Brennecke]{Ferdinand Brennecke}

\address{Institut f\"ur Quantenelektronik \\
ETH Z\"urich\\
CH-8093 Z\"urich 
Switzerland}

\email{brennecke@phys.ethz.ch}

\author[D\"utsch]{Michael D\"utsch}

\address{Institut f\"ur Theoretische Physik\\ 
Universit\"at Z\"urich\\
CH-8057 Z\"urich 
Switzerland}

\email{duetsch@physik.unizh.ch}


\date{January 1, 2004}

\begin{abstract}
In perturbative quantum field theory the maintenance of classical symmetries is  quite often
investigated by means of algebraic renormalization, which is based on the Quantum Action Principle.
We formulate and prove this principle in a new framework, in causal perturbation theory
with localized interactions.
Throughout this work a universal formulation of symmetries is used: the Master Ward Identity. 
\end{abstract}

\maketitle
\section{Introduction}
The main problem in perturbative renormalization is to prove that symmetries of the underlying classical theory 
can be maintained in the process of renormalization. In traditional renormalization theory this is done by 
'algebraic renormalization' \cite{Piguet:1995er}.
This method relies on the '{\it Quantum Action Principle}' (QAP), which is due to Lowenstein 
\cite{Lowenstein:1971jk} and Lam \cite{Lam:1972mb}. This principle states that the most general violation of an identity expressing a
relevant symmetry ('Ward identity') can be expressed by the insertion of a local field with 
appropriately bounded mass dimension. Proceeding in a proper field formalism\footnote{By 'proper field formalism'
we mean the description of a perturbative QFT in terms of the generating functional of the 1-particle irreducible diagrams.}
by induction on the order of $\hbar$, this knowledge about the structure of violations of Ward identities and often 
cohomological results are used to remove these violations by finite renormalizations.  For example, this method has been
used to prove BRST-symmetry of Yang-Mills gauge theories 
\cite{Becchi:1974md,Becchi:1975nq,Tyutin:1975qk,Henneaux:1992ig,Barnich:2000zw}.

Traditionally, algebraic renormalization is formulated in terms of a renormalization method in which the interaction is
not localized (i.e. $S_\mathrm{int}=\int dx\,\mathcal{L}_\mathrm{int}(x)\,$, where $\mathcal{L}_\mathrm{int}$ 
is a polynomial in the basic fields with constant coefficients), for example the BPHZ momentum space subtraction procedure
\cite{Zimmermann:1969,Lowenstein:1971jk,Lam:1972mb} or the pole subtractions 
of dimensionally regularized integrals \cite{Breitenlohner:1977hr}.
In \cite{Piguet-Rouet} it is pointed out (without proof) that the QAP is a general theorem in perturbative QFT
for non-localized interactions, i.e.~it holds in any renormalization scheme.\footnote{Causal perturbation theory, 
with the adiabatic limit carried out, is included in that statement.}

However, for the generalization of perturbative QFT to general globally hyperbolic {\it curved spacetimes},
it is advantageous to work with {\it localized interactions} (i.e. $S_\mathrm{int}=\int dx\,
\sum_{n\geq 1} (g(x))^n\,\mathcal{L}_{\mathrm{int},n}(x)\,$, 
where $g$ is a test function with compact support) and to use a renormalization method 
which proceeds in configuration space and in which the locality and causality
of perturbative QFT is clearly visible \cite{Brunetti:1999,Hollands:2002,Hollands:2004yh}.
It is {\it causal perturbation theory} (CPT) \cite{Bogoliubov:1959,Epstein:1973gw,Duetsch:2004dd} 
which is distinguished by these criteria.

Since it is the framework of {\it algebraic QFT} \cite{Haag} in which the problems 
specific for curved spacetimes (which mainly rely on the absence of translation invariance) can best be treated,
our main goal is the perturbative construction of the net of local algebras of interacting fields ('perturbative
algebraic QFT'). Using the formulation of 
causality in CPT, it was possible to show that for this construction it is sufficient 
to work with {\it localized} interactions \cite{Brunetti:1999,Duetsch:2000de}. 
Hence, a main argument against localized interactions, namely that  
a space or time dependence of the coupling constants has not been observed in experiments, 
does not concern perturbative algebraic QFT. 
Because of the localization of the interactions, 
the construction of the local algebras of interacting fields is not plagued by infrared divergences, the latter
appear only in the construction of physical states.

Due to these facts it is desirable to transfer the techniques of algebraic renormalization to CPT, that is
to formulate the $\hbar$-expansion, a proper field formalism and the QAP in the 
framework of CPT. For the $\hbar$-expansion the difficulty is that CPT is a construction of the perturbation series
by induction on the coupling constant, a problem solved in \cite{Duetsch:2000nh,Duetsch:2000de}.  
A formulation of the QAP in the 
framework of CPT has partially been given in \cite{Duetsch:2000nh} and in \cite{Pinter}; but for symmetries relying on a variation 
of the fields (as e.g.~BRST-symmetry) an appropriate formulation and a proof were missing up to the appearance of the paper
\cite{qap}. In the latter, also a proper field formalism and algebraic renormalization are developed
in the framework of CPT.

In this paper we concisely review main results of that work \cite{qap}, putting the focus on the QAP. To be closer to
the conventional treatment of perturbative QFT in Minkowski space and to simplify the formalism, we work with the Wightman 2-point 
function instead of a Hadamard function.\footnote{In \cite{qap} smoothness in the mass $m$ is required for $m\geq 0$
which excludes the  Wightman 2-point function.} Compared with \cite{qap}, 
we formulate some topics alternatively, in particular we introduce the proper field formalism without 
using arguments relying on Wick's theorem and the corresponding diagrammatic interpretation.
In addition we prove a somewhat stronger version of the QAP.

The validity of the QAP is very general. Therefore, we investigate a universal formulation of Ward identities: the {\it Master 
Ward Identity} (MWI) \cite{Duetsch:2001sw,Duetsch:2002yp}.
This identity can be derived in the framework of classical field theory simply from the fact 
that classical fields can be multiplied pointwise. Since this is impossible for quantum fields
(due to their distributional character), the MWI is a highly non-trivial renormalization condition, which cannot 
be fulfilled in general, the well known anomalies of perturbative QFT are the obstructions. 

\section{The off-shell Master Ward Identity in classical field theory}

For algebraic renormalization it is of crucial importance that the considered Ward identities hold true in classical field theory.
Therefore, in this section, we derive the off-shell MWI in the classical framework. The formalism  
of classical field theory, which we are going to introduce, will be used also
in perturbative QFT, since the latter will be obtained by deformation of the classical Poisson algebra (Sect.~3) 
\cite{Duetsch:2000nh,Duetsch:2000de,Duetsch:2002yp,Duetsch:2004dd}.

For simplicity we study the model of a real scalar field $\varphi$ on $d$ dimensional 
Minkowski space $\MM$, $d>2$. The field $\varphi$ and partial derivatives $\d^a\varphi$
($a\in\NN_0^d$) are evaluation functionals on the configuration space 
$\mathcal{C}\equiv\mathcal{C}^{\infty}(\MM,\RR)\,$:
$(\d^a\varphi) (x)(h)=\d^ah(x)$. Let $\mathcal{F}$ 
be the space of all functionals  
\beq
F(\varphi)\,:\,\mathcal{C}\longrightarrow\CC\,,\,\quad
F(\varphi)(h)=F(h)\ ,\label{functionals}
\eeq
which are localized polynomials in $\varphi$:
\begin{equation}
  F(\varphi)=\sum_{n=0}^N\int\! dx_1\ldots dx_n\,\varphi(x_1)
  \cdots\varphi(x_n)f_n(x_1,\ldots,x_n)\ ,
  \label{F(phi)} 
\end{equation}
where $N<\infty$ and the $f_n$'s are $\CC$-valued 
distributions with compact support, which are symmetric under 
permutations of the arguments and whose wave front 
sets satisfy the condition 
\begin{equation}
  \mathrm{WF}(f_n)\cap \bigl(\MM^n\times
(\overline{V}_+^{\,n}\cup \overline{V}_-^{\,n})\bigr)
   =\emptyset \label{WF}
\end{equation}
and $f_0\in\CC$. ($\overline{V}_\pm$ denotes the closure of the forward/backward light-cone.)
Endowed with the {\it classical product} $(F_1\cdot F_2)(h):=F_1(h)\cdot F_2(h)$,
the space $\mathcal{F}$ becomes a commutative algebra. By the support of a functional 
$F\in\mathcal{F}$ we mean the support of $\tfrac{\delta F}{\delta\varphi}$.

The space of {\it local functionals} $\mathcal{F}_{\rm loc}\subset \mathcal{F}$ is defined as
\begin{equation}
\mathcal{F}_{\rm loc}\=d
  \Big\{\int\! dx\,\sum_{i=1}^N A_i(x)h_i(x) \equiv \sum_{i=1}^{N} 
  A_{i}(h_{i})\,|\, A_i\in \mathcal{P}\, ,\,h_i\in \mathcal{D}(\MM)\Big\}\ ,\label{F_loc}
\end{equation}
where $\mathcal{P}$ is the linear space of all polynomials of the field $\varphi$ and its 
partial derivatives:
\beq
\mathcal{P}:=\bigvee \Big\{\d^a\varphi\,|\,a\in\NN_0^d\Big\}\ .
\eeq

We consider action functionals of the form 
$S_{\mathrm{tot}}= S_0+\lambda\,S$ where $S_0 \defi \int \! dx \frac{1}{2}(\del_\mu \varphi 
\del^\mu \varphi -m^2 \varphi^2)$ is the free action, $\lambda$ a real parameter 
and $S\in \mathcal{F}$ some compactly supported interaction,
which may be {\it non-local}. The retarded Green 
function $\Delta^\mathrm{ret}_{S_\mathrm{tot}}$
corresponding to the action $S_\mathrm{tot}$, is defined by
\begin{equation}\label{def:green}
\int\! dy\, \Delta^\mathrm{ret}_{S_\mathrm{tot}}(x, y)\frac{\delta^2 S_\mathrm{tot}}{\delta \varphi(y) 
\delta \varphi(z)}=\delta(x-z)=\int \!dy\, \frac{\delta^2 S_\mathrm{tot}}{\delta \varphi(x) 
\delta \varphi(y)}\Delta^\mathrm{ret}_{S_\mathrm{tot}}(y, z)
\end{equation}
and $\Delta^\mathrm{ret}_{S_\mathrm{tot}}(x,y)=0$ for $x$ sufficiently early. In the following we consider only 
actions $S_\mathrm{tot}$ for which the retarded Green function exists and is unique
in the sense of formal power series in $\lambda$.

To introduce the perturbative expansion around the free theory and to define the Peierls bracket,
we define retarded wave operators which 
map solutions of the free theory to solutions of the interacting theory \cite{Duetsch:2002yp}. 
However, we define them as maps on the space $\mathcal{C}$ of all field 
configurations ('\emph{off-shell formalism}') and not only on the space of free solutions:

\begin{defn}\label{def:waveop}
A retarded wave operator is a family of maps $(r_{S_0+S, S_0})_{S\in \mathcal{F}}$ 
from $\mathcal{C}$ into itself with the properties
\begin{enumerate}
\item $r_{S_0+S, S_0}(f)(x)=f(x)$ for $x$ sufficiently early
\item $\frac{\delta(S_0+S)}{\delta \varphi}\circ r_{S_0+S, S_0}=\frac{\delta S_0}{\delta \varphi}$.
\end{enumerate}
\end{defn}
The following Lemma is proved in \cite{qap}.
\begin{lem}
The retarded wave operator $(r_{S_0+ S, S_0})_{S\in \mathcal{F}}$ exists and is unique 
and invertible in the sense of formal power series in the interaction $S$.
\end{lem}

Motivated by the interaction picture known from QFT, we introduce retarded fields:
the classical retarded field to the interaction $S$
and corresponding to the functional $F\in \mathcal{F}$ is defined by
\begin{equation}
F^{\mathrm{cl}}_{S}\defi F\circ r_{S_0+ S, S_0}: \mathcal{C}\longrightarrow \CC.\label{cl-ret-field}
\end{equation}
The crucial  factorization property,
\begin{equation}\label{interfactor}
(F\cdot G)^{\mathrm{cl}}_{S}=F^{\mathrm{cl}}_{S}\cdot G^{\mathrm{cl}}_{S}\ ,
\end{equation}
cannot be maintained in the process of quantization, because quantum fields are distributions.
This is why many proofs of symmetries in classical field theory do not apply to QFT (cf.~Sect.~5).

The perturbative expansion around the free theory is defined by expanding the retarded fields with 
respect to the interaction. The coefficients are given by the classical retarded product $R_\mathrm{cl}$ 
\cite{Duetsch:2002yp}:
\begin{equation}\label{retprod}
R_\mathrm{cl}:\mathbb{T}\mathcal{F}\otimes \mathcal{F} \rightarrow \mathcal{F}\ ,
\quad R_\mathrm{cl}(S^{\otimes n}, F)\defi 
\frac{d^n}{d\lambda^n}\Big|_{\lambda=0}F\circ r_{S_0+\lambda S, S_0}\ ,
\end{equation}
where $\mathbb{T}\mathcal{V}\defi \CC\oplus\bigoplus_{n=1}^\infty \mathcal{V}^{\otimes n}$ denotes the 
tensor algebra corresponding to some vector space $\mathcal{V}$. For non-diagonal entries,
$R_\mathrm{cl}(\otimes_{j=1}^n S_j,F)$ is determined by linearity and symmetry under permutations of $S_1,...,S_n$.
Interacting fields can then be written as
\begin{equation}
F^{\mathrm{cl}}_{S}\simeq\sum_{n=0}^\infty \frac{1}{n!}R_\mathrm{cl}(S^{\otimes n}, F)\equiv R_\mathrm{cl}(e_\otimes^S, F)\ .
\end{equation}
The r.h.s. of $\simeq$ is interpreted as a {\it formal power series} (i.e.~we do not care about convergence of the series). 

By means of the retarded wave operator one can define an off-shell version \cite{qap} of the Peierls bracket
associated to the action $S$ \cite{Peierls:1952}, $\{\cdot,\cdot\}_S\,:\,\mathcal{F}\otimes \mathcal{F} \rightarrow \mathcal{F}$,
and one verifies that this is indeed a Poisson bracket, i.e.~that  $\{\cdot,\cdot\}_S$ is linear, antisymmetric 
and satisfies the Leibniz rule and the Jacobi identity \cite{Duetsch:2002yp,qap}.
\medskip

Following \cite{qap}, we are now going to derive the classical 
off-shell MWI from the factorization \eqref{interfactor} and the definition
of the retarded wave operators. Let $\mathcal{J}$ be the ideal generated by the free field equation,
\beq
\mathcal{J}\defi \Big\{\sum_{n=1}^{N} \int\! dx_1\ldots dx_n \,\varphi(x_1) \cdots \varphi(x_{n-1})
\frac{\delta S_0}{\delta \varphi(x_n)} f_{n}(x_1, \ldots, x_n)\Big\}\subset \mathcal{F}\ ,\nonumber
\eeq
with $N<\infty$ and the $f_{n}$'s being defined as in \eqref{F(phi)}. Obviously, every 
$A\in \mathcal{J}$ can be written as
\begin{equation}
A\defi \int\! dx\, Q(x) \ffgx\ ,
\end{equation}
where $Q$ may be non-local. Given 
$A\in \mathcal{J}$ we introduce a corresponding derivation \cite{Duetsch:2002yp}
\begin{equation}\label{defdeltaA}
\dA \defi \int\! dx\, Q(x)\frac{\delta}{\delta \varphi(x)}.
\end{equation}
Notice $F(\varphi+Q)-F(\varphi)=\delta_A F+\mathcal{O}(Q^2)$ (for $F\in\mathcal{F}$)
that is, $\delta_A F$ can be interpreted as the variation of $F$ under the infinitesimal 
field transformation $\varphi(x)\mapsto\varphi(x)+Q(x)$. 
From the definition of the retarded wave operators Def.~\ref{def:waveop} we obtain
\begin{eqnarray}
(A+\delta_A S)\circ r_{S_0+S, S_0}&=&\int\! dx\, Q(x)\circ r_{S_0+S, S_0}
\frac{\delta(S_0+S)}{\delta \varphi(x)}\circ r_{S_0+S, S_0}\nonumber\\
&=& \int\! dx\, Q(x)\circ r_{S_0+S, S_0}\frac{\delta S_0}{\delta \varphi(x)}\ .\label{klMWI}
\end{eqnarray}
In terms of the perturbative expansion this relation reads
\begin{equation}\label{klMWIIdeal}
R_{\mathrm{cl}}(e_\otimes^S, A+\delta_A S)=\int\! dx\,
R_{\mathrm{cl}}\big(e_\otimes^S, Q(x)\big)\frac{\delta S_0}{\delta
\varphi(x)}\in \mathcal{J}.
\end{equation}
This is the MWI written in the off-shell formalism. 
When restricted to the solutions of the free field equation, the right-hand side vanishes and we 
obtain the on-shell version of the MWI, as it was derived in \cite{Duetsch:2002yp}.
For the simplest case $Q=1$ the MWI 
reduces to the off-shell version of the (interacting) field equation
\begin{equation}\label{klfeldgloff}
R_{\mathrm{cl}}\Big(e_\otimes^S, \frac{\delta(S_0+S)}{\delta \varphi(x)}\Big)=\frac{\delta S_0}{\delta \varphi(x)}\ .
\end{equation} 

\section{Causal perturbation theory}

Following \cite{Duetsch:2004dd}, we quantize
perturbative classical fields by deforming the underlying free theory as a 
function of $\hbar$: we replace 
$\mathcal{F}$ by $\mathcal{F}[[\hbar]]$ (i.e.~all functionals are formal power 
series in $\hbar$) and deform the classical product into the 
$\star$-product, $\star\,:\, \mathcal{F}\times \mathcal{F}\rightarrow \mathcal{F}$ (for
simplicity we write $ \mathcal{F}$ for $\mathcal{F}[[\hbar]]$):
\begin{gather}
  (F\star G)(\varphi)\=d
  \sum_{n=0}^{\infty}\frac{\hbar^n}{n!}
  \int\! dx_1\ldots dx_n dy_1\ldots dy_n 
  \frac{\delta^n F}{\delta\varphi(x_1)\cdots\delta\varphi(x_n)}\notag\\
  \cdot \prod_{i=1}^n \Delta^+_m(x_i-y_i) 
  \frac{\delta^n G}{\delta\varphi(y_1)\cdots\delta\varphi(y_n)}\ .
  \label{*-product}
\end{gather}
The $\star$-product is still associative but non-commutative.

In contrast to the classical retarded field $F_S^\mathrm{cl}$ \eqref{cl-ret-field}, 
one assumes in perturbative QFT that the interaction $S$ and the 
field $F$ are {\it local} functionals. For an interacting quantum field $F_S$  one makes the 
ansatz of a \emph{formal power series} in the interaction $S$:
\begin{equation}
F_{S}=\sum_{n=0}^\infty\frac{1}{n!}R_{n,1}
\bigl(S^{\otimes n},F \bigr)\equiv  
R(e_\otimes^{S},F)\ .
\label{ansatz:intfield}
\end{equation}
The 'retarded product' $R_{n,1}$ is a {\bf linear} map, 
from $\mathcal{F}_{\rm loc}^{\otimes n}\otimes \mathcal{F}_{\rm loc}$ into $\mathcal{F}$ which is 
{\bf symmetric in the first $n$ variables}. We interpret $R(A_1(x_1),...;A_n(x_n))\,$,
$A_1,...,A_n\in \mathcal{P}$, as $\mathcal{F}$-valued distributions on $\mathcal{D}(\MM^n)$,
which are defined by: $\int dx\, h(x)\,R(...,A(x),...):=R(...\otimes A(h)\otimes...)\>\>
\forall h\in \mathcal{D}(\MM)$. 

Since the retarded products depend only on the functionals (and not on how the 
latter are written as smeared fields (\ref{F_loc})),
they must satisfy the {\bf Action Ward Identity (AWI)} \cite{Duetsch:2004dd,Stora:2002,Stora:2006}:
\begin{equation}
  \d^x_{\mu}R_{n-1,1}(\ldots A_k(x)\ldots)= 
   R_{n-1,1}(\ldots,\d_{\mu}A_k(x),\ldots)\ .
\end{equation}

Interacting fields are defined by the following axioms \cite{Duetsch:2004dd},
which are motivated by their validity in classical field theory. The
{\bf basic axioms} are the initial condition $R_{0,1}(1,F)=F$ and
\begin{description}
\item[Causality] $F_{G+H} = F_{G}\>\>$ if $\>\>\supp(\tfrac{\delta F}{\delta\varphi})\cap 
(\supp(\tfrac{\delta H}{\delta\varphi})+\bar V_+)=\emptyset$\ y;
\item[GLZ Relation] 
$F_{G}\star H_{G}-H_{G}\star F_{G} = \frac{d}{d\lambda}\Big|_{\lambda=0}\, 
\left(F_{G+\lambda H}- H_{G+\lambda F}\right) \ .$
\end{description}
Using only these requirements, the retarded products $R_{n,1}$ can be
constructed by induction on $n$ (cf.~\cite{Steinmann:1971}). However, in each
inductive step one is free to add a local functional,
which corresponds to the usual renormalization ambiguity.  This
ambiguity is reduced by imposing {\bf renormalization conditions} as
further axioms, see below.

Mostly, perturbative QFT is formulated in terms of the {\bf time ordered product} ('$T$-product') $T\,:\,
\TT \mathcal{F}_{\rm loc}\rightarrow \mathcal{F}$, which is a {\bf linear} and 
{\bf totally symmetric} map. Compared with the $R$-product, the $T$-product has 
the advantage of being totally symmetric and the disadvantage that its classical limit does not 
exist \cite{Duetsch:2000nh}. $R$- and $T$-products are related by Bogoliubov's formula: 
\beq
 R\bigl(e_\otimes^S\,,\,F\bigr)=\frac{\hbar}{i}\,\mathbf{S}(S)^{-1}
\star  \frac{d}{d\tau}\Big\vert_{\tau=0}\, \mathbf{S}(S+\tau F)\ ,\label{R1-T}
\eeq
where
\beq
 \mathbf{S}(S)\equiv T\bigl(e_\otimes^{iS/\hbar}\bigr)\equiv
\sum_{n=0}^\infty\frac{i^n}{n!\hbar^n}\,T_n(S^{\otimes n})\ .
\eeq
The basic axioms for retarded products translate into 
the following basic axioms for $T$-products: the initial conditions $T_0(1)=1$, $T_1(F)=F$ 
and {\bf causal factorization}:
\begin{gather}
T_n(A_1(x_1),...,A_n(x_n))=\notag\\
T_k(A_1(x_1),...,A_k(x_k))\star T_{n-k}(A_{k+1}(x_{k+1}),...,A_n(x_n))\label{causfac}
\end{gather}
if $\{x_1,...,x_k\}\cap (\{x_{k+1},...,x_n\}+\bar V_-) =\emptyset$. 
There is no axiom corresponding to the GLZ Relation. The latter can be interpreted 
as 'integrability condition' for the 'vector potential' $R\bigl(
e_\otimes^S\,,\,F\bigr)$, that is it ensures the existence of the
'potential' $\mathbf{S}(S)$  fulfilling (\ref{R1-T}); for details see \cite{BDF} and
Proposition 2 in \cite{DF:QED}.

For this paper the following {\bf renormalization conditions} are relevant (besides the MWI).
\begin{description}
\item[Translation Invariance] The group $(\RR^d,+)$ of space and time translations has an obvious
automorphic action $\beta$ on $\mathcal{F}$, which is determined by $\beta_a\varphi(x)=\varphi(x+a)\ ,
\,\, a\in\RR^d$. We require
\beq
\beta_a\,\mathbf{S}(S)=\mathbf{S}(\beta_a S)\ ,\quad\forall a\in\RR^d\ .\label{transinv}
\eeq
\item[Field Independence] $\frac{\delta T}{\delta\varphi(x)}=0\,$. This axiom implies the causal Wick expansion of 
\cite{Epstein:1973gw} as follows \cite{Duetsch:2004dd}: since $T(\otimes_{j=1}^n F_j)\in\mathcal{F}$ is
polynomial in $\varphi$, it has a finite Taylor expansion in $\varphi$. By using Field Independence, this
expansion can be written as
\begin{gather}
T_n(A_1(x_1), \cdots , A_n(x_n))=
\sum_{l_1, \ldots, l_n}\frac{1}{l_1!\cdots l_n!} \notag\\
 \cdot T_n\Big(\cdots ,\sum_{a_{i1}\ldots a_{il_i}}\frac{\del^{l_i}A_i}{\del(\del^{a_{i1}} \varphi)
\cdots \del(\del^{a_{il_i}}\varphi)}(x_i),\cdots\Big)\Big\vert_{\varphi=0}
\prod_{i=1}^{n}\prod_{j_i=1}^{l_i}\del^{a_{ij_i}}\varphi(x_i)\label{causWick}
\end{gather}
with multi-indices $a_{ij_i}\in \mathbb{N}^d_0$.
\item[Scaling] This requirement uses the \textit{mass dimension of a monomial} in 
$\mathcal{P}$, which is defined by the conditions
\begin{equation}
  \mathrm{dim}(\d^a\varphi)=\frac{d-2}{2}+|a|\quad
\mathrm{and}\quad \mathrm{dim}(A_1A_2)=\mathrm{dim}(A_1)+
\mathrm{dim}(A_2)\label{UV-dim}
\end{equation}
for all monomials $A_1,A_2\in \mathcal{P}$. The mass dimension of a
\textit{polynomial} in $\mathcal{P}$ is the maximum of the mass dimensions
of the contributing monomials. We denote by $\mathcal{P}_{\text{hom}}$ the set
of all field polynomials which are homogeneous in the mass dimension.

The axiom {\bf Scaling Degree}
requires  that 'renormalization may not make the interacting fields 
more singular' (in the UV-region). Usually this is formulated in terms of 
Steinmann's \textit{scaling degree} \cite{Steinmann:1971}:
 \begin{equation}
{\rm sd}(f)\=d {\rm inf}\{\delta\in \RR\>|\>\lim_{\rho\downarrow 0}
\rho^\delta f(\rho x)=0\},\quad f\in \mathcal{D}'(\RR^k)
\>\>\mathrm{or}\>\> f\in \mathcal{D}'(\RR^k\setminus \{0\}).\label{sd}
\end{equation}
Namely, one requires
\beq
{\rm sd}\Bigl(T(A_1,...,A_n)\vert_{\varphi=0}(x_1-x_n,...)\Bigr)\leq\sum_{j=1}^n
{\rm dim}(A_j)\ ,\quad\forall A_j\in \mathcal{P}_{\rm hom}\ ,\label{axiom-sd}
\eeq
where Translation Invariance is assumed. Notice that this condition restricts \textit{all}
coefficients in the causal Wick expansion \eqref{causWick}.
\end{description}
In the inductive construction of the sequence $(R_{n-1,1})_{n\in \NN}$ or
$(T_n)_{n\in \NN}$, respectively, the problem of renormalization appears as the 
extension of the coefficients in the causal Wick expansion (which are
$\CC[[\hbar]]$-valued distributions) 
from $\mathcal{D}(\RR^{d(n-1)}\setminus\{ 0\})$ to $\mathcal{D}(\RR^{d(n-1)})$.
This extension has to be done in the sense of formal power series 
in $\hbar$, that is individually in each order in $\hbar$. With that it holds
\beq
\lim_{\hbar\to 0} R=\Rcl\ .\label{R->R_cl}
\eeq

In \cite{Duetsch:2004dd} it is shown that there {\bf exists} a $T$-product which fulfills all 
axioms. The {\bf non-uniqueness} of solutions is characterized by the 'Main Theorem'; 
for a complete version see \cite{Duetsch:2004dd}.

\section{Proper vertices}

A main motivation for introducing proper vertices is to select that part of a 
$T$-product for which renormalization is non-trivial (cf. \cite{Jona-Lasinio:1964}). 
This is the contribution of
all 1-particle-irreducible (1PI) subdiagrams. This selection can be done as follows: 
first one eliminates all disconnected diagrams. Then, one interprets each connected diagram  
as tree diagram with {\it non-local} vertices ('proper vertices') given by the     
1PI-subdiagrams. The proper vertices can be interpreted as the 'quantum part' of the Feynman diagrams.
Since renormalization is unique and trivial for tree diagrams, Ward identities can
equivalently be formulated in terms of proper vertices (Sect.~5.1).

Essentially we follow this procedure, however, we avoid to argue in terms of diagrams, i.e.~to 
use Wick's Theorem. It has been shown in \cite{qap}
that with our definition (\ref{Gamma}) of the vertex functional $\Gamma$ 
the 'proper interaction' $\Gamma(\ets)$ corresponds to the sum of all 1PI-diagrams of $T(e_\otimes^{iS/\hbar})$.

The connected part $T^c$ of a time-ordered $T$ can be defined recursively by \cite{Duetsch:2000nh}
\beq
T_n^c(\otimes_{j=1}^n F_j)\=d T_n(\otimes_{j=1}^n F_j)-
\sum_{|P|\geq 2}\prod_{J\in P}T_{|J|}^c(\otimes_{j\in J} F_j)\ .\label{T^c}
\eeq
It follows that $T$ and $T^c$ are related by the linked cluster theorem:
\beq
T(e_\otimes^{iF})={\rm exp}_\bullet (T^c(e_\otimes^{iF}))\ ,\label{linked-cluster}
\eeq
where ${\rm exp}_\bullet$ denotes the exponential function with respect to the classical product.

For $F\in \mathcal{F}_{\mathrm{loc}}$ the connected tree part $T^c_{{\rm tree},n}(F^{\otimes n})$
can be defined as follows \cite{Duetsch:2000nh}: since $T_n^c=\mathcal{O}(\hbar^{n-1})$ , the limit
\beq
\hbar^{-(n-1)}\,T^c_{{\rm tree},n}\=d\lim_{\hbar\to 0}\hbar^{-(n-1)}\,T^c_n \label{T^c_tree}
\eeq
exists. This definition reflects the well known statements that $T^c_{\rm tree}$ is the 
'classical part' of $T^c$ and that connected loop diagrams are of higher orders in $\hbar$.

Since proper vertices are non-local, we need the connected tree part $T^c_{\rm tree}(\otimes_{j=1}^n F_j)$ 
for non-local entries $F_j\in\mathcal{F}$. This can be defined recursively \cite{qap}:
\begin{gather} 
T^c_{\rm tree}(\otimes_{j=1}^{n+1} F_j)=\sum_{k=1}^n\int\! dx_1...dx_k\, 
dy_1...dy_k\,\frac{\delta^k F_{n+1}}{\delta\varphi(x_1)...\delta\varphi(x_k)}\,\notag\\ 
\cdot\prod_{j=1}^k \Delta^F_m(x_j-y_j)\,\frac{1}{k!}\,\sum_{I_1\sqcup...\sqcup I_k=\{1,...,n\}} 
\frac{\delta}{\delta\varphi(y_1)} T^c_{\rm tree}(\otimes_{j\in I_1} F_j)\cdot...\notag\\ 
\hspace{4.5cm}\cdot\frac{\delta}{\delta\varphi(y_k)} T^c_{\rm tree}(\otimes_{j\in I_k} F_j)\ , 
\label{conntree:recursion} 
\end{gather} 
where $I_j\not=\emptyset\,\,\forall j\,$,   
$\sqcup$ means the disjoint union and $\Delta^F_m$ is the Feynman propagator for mass $m$. 
(Note that in the sum over $I_1,...,I_k$ the order of $I_1,...,I_k$ 
is distinguished and, hence, there is a factor $\frac{1}{k!}$.) 
For local entries the two definitions (\ref{T^c_tree}) and (\ref{conntree:recursion})
of $T^c_{\rm tree}$ agree, as explained in \cite{qap}.

The 'vertex functional' $\Gamma$ is defined by the following proposition \cite{qap}: 
\begin{prop}\label{prop:gamma} 
There exists a {\bf totally symmetric} and {\bf linear} map 
\beq 
\Gamma\> :\> \TT \mathcal{F}_{\rm loc}\rightarrow \mathcal{F} 
\label{gamma} 
\eeq 
which is uniquely determined by 
\beq 
T^c(e_\otimes^{iS/\hbar})= 
T^c_{\rm tree}\Bigl(e_\otimes^{i\Gamma(e_\otimes^S)/\hbar}\Bigr)\ . 
\label{Gamma} 
\eeq 
\end{prop} 
To zeroth and first order in $S$ we obtain 
\beq 
\Gamma(1)=0\ ,\quad \Gamma(S)=S\ .\label{Gamma:0,1} 
\eeq 
Since $T^c$, $T^c_{\rm tree}$ and $\Gamma$ are linear and totally symmetric,
the defining relation (\ref{Gamma}) implies 
\beq 
T^c(e_\otimes^{iS/\hbar}\otimes F)= 
T^c_{\rm tree}\Bigl(e_\otimes^{i\Gamma(e_\otimes^S)/\hbar} 
\otimes \Gamma(e_\otimes^S\otimes F)\Bigr)\ . 
\eeq 

To prove the proposition, one constructs $\Gamma(\otimes_{j=1}^n F_j)$ by induction on $n$,   
using (\ref{Gamma}) and the requirements total symmetry and linearity:
\beq 
\Gamma(\otimes_{j=1}^n F_j)=(i/\hbar)^{n-1}\,T^c(\otimes_{j=1}^n F_j)-\sum_{|P|\geq 2} 
(i/\hbar)^{|P|-1}\,T^c_{\rm tree}\Bigl(\bigotimes_{J\in P} 
\Gamma(\otimes_{j\in J} F_j)\Bigr)\ ,\label{Gamma:recursion} 
\eeq 
where $P$ is a partition of $\{1,...,n\}$ in $|P|$ subsets $J$. 

From this recursion relation and from $T_n^c -T^c_{{\rm tree},n}=\mathcal{O}(\hbar^{n})$ 
we inductively conclude 
\beq\label{properinter} 
\Gamma(e_\otimes^S)=S+\mathcal{O}(\hbar)\ ,\quad \Gamma(e_\otimes^S 
\otimes F)=F+\mathcal{O}(\hbar)\quad\mathrm{if}\quad F,S\sim\hbar^0\ .
\eeq 
Motivated by this relation and (\ref{Gamma}) we call $\Gamma(e_\otimes^S)$ the 'proper 
interaction' corresponding to the classical interaction $S$.

The validity of renormalization conditions 
for $T$ implies corresponding properties of $\Gamma$, as worked out in \cite{qap}.

Analogously to the conventions for $R$- and $T$-products we sometimes write\\ 
$\int\! dx\, g(x)\, \Gamma(A(x)\otimes F_2...)$ for $\Gamma(A(g)\otimes F_2...)$
($A\in\mathcal{P}\,,\,g\in\mathcal{D}(\MM)$).
Since $\Gamma$ depends only on the {\it functionals}, 
it fulfills the AWI: $\d^\mu_x\Gamma(A(x)\otimes F_2...)= 
\Gamma(\d^\mu A(x)\otimes F_2...)$. 

\section{The Quantum Action Principle}

\subsection{Formulation of the Master Ward Identity in terms of proper vertices}

The classical MWI was derived for arbitrary interaction $S\in \mathcal{F}$ and 
arbitrary $A\in \mathcal{J}$. For {\it local} functionals $S\in \mathcal{F}_{\mathrm{loc}}$ and
\beq
A=\int\! dx\, h(x)\, Q(x)\frac{\delta S_0}{\delta \varphi(x)}\in\mathcal{J}\cap\mathcal{F}_{\mathrm{loc}}\ ,
\quad  h\in \mathcal{D}(\MM)\ ,\quad Q\in \mathcal{P}\ ,\label{A:loc}
\eeq
it can be transferred formally into perturbative QFT (by the replacement $R_\mathrm{cl}\rightarrow R$),
where it serves as an additional, highly non-trivial renormalization condition:
\begin{equation}\label{MWI:R}
R\big(\ets, A+\delta_A S\big)=\int\! dy\, h(y) R(\ets, Q(y))\frac{\delta S_0}{\delta \varphi(y)}\ .
\end{equation}
Since the MWI holds true in classical field theory (i.e.~for connected tree diagrams, see below) it is possible
to express this renormalization condition in terms of the 'quantum part' (described by the loop diagrams)
- that is in terms of proper vertices. We do this in several steps:

\paragraph{Proof of the MWI for $T^c_\mathrm{tree}$ (connected tree diagrams).} Since this is an alternative 
formulation of the {\it classical} MWI, we still include {\it non-local functionals} $S\in \mathcal{F}$,
$A=\int\! dx\, Q(x)\frac{\delta S_0}{\delta \varphi(x)}\in\mathcal{J}$, as in Sect.~2.
The classical field equation \eqref{klfeldgloff} can be expressed in terms of $T^c_\mathrm{tree}$:
\beq
T^c_\mathrm{tree}\Bigl(e_\otimes^{iS/\hbar}\otimes\frac{\delta (S_0+S)}{\delta\varphi(x)}\Bigr)
=\frac{\delta S_0}{\delta\varphi(x)}\ .
\eeq
The only difference between $R_\mathrm{cl}$ and $T^c_\mathrm{tree}$ is that the retarded propagator
$\Delta^\mathrm{ret}(y)(\not=\Delta^\mathrm{ret}(-y))$ is replaced by the Feynman propagator
$\Delta^F(y)(=\Delta^F(-y))$, the combinatorics of the diagrams remains the same. Hence, the factorization 
of classical fields \eqref{interfactor},
\beq
R_\mathrm{cl}\big(\ets,F\cdot G\big)=
R_\mathrm{cl}\big(\ets,F\big)\cdot R_\mathrm{cl}\big(\ets,G\big)
\eeq
holds true also for  $T^c_\mathrm{tree}$:
\beq
T^c_{\rm tree}\Bigl(e_\otimes^{iS/\hbar}\otimes F G\Bigr)=
T^c_{\rm tree}\Bigl(e_\otimes^{iS/\hbar}\otimes F\Bigr)\cdot
T^c_{\rm tree}\Bigl(e_\otimes^{iS/\hbar}\otimes G\Bigr)\ .\label{factorization}
\eeq
We now multiply the field equation for $T^c_{\rm tree}$ with $T^c_{\rm tree}(e_\otimes^{iS/\hbar}\otimes Q(x))$ 
and integrate over $x$. This yields the MWI for $T^c_\mathrm{tree}$:
\beq
T^c_{\rm tree}\Bigl(e_\otimes^{iS/\hbar}\otimes (A+\delta_A S)\Bigr)=
\int \!dx\,
T^c_{\rm tree}\Bigl(e_\otimes^{iS/\hbar}\otimes Q(x)\Bigr)\cdot
\frac{\delta S_0}{\delta\varphi(x)}\ .\label{MWI-tree}
\eeq

\paragraph{Translation of the (quantum) MWI from $R$ into $T^c$.} Using Bogoliubov's formula \eqref{R1-T}
and the identity
\beq
(F\star G)\cdot\frac{\delta S_0}{\delta \varphi}=F\star \big(G \cdot \frac{\delta S_0}{\delta\varphi}\big)
\qquad \forall F, G\in \mathcal{F}\label{hilfsid}
\eeq
(which relies on $(\square+m^2)\Delta^+_m=0$), the MWI in terms of $R$-products \eqref{MWI:R}
can be translated into $T$-products:
\begin{equation}\label{MWI:T} 
T\big(e_\otimes^{iS/\hbar}\otimes (A+\delta_A S)\big)=\int\! dy\, h(y) 
\,T(e_\otimes^{iS/\hbar}\otimes Q(y))\frac{\delta S_0}{\delta \varphi(y)}\ ,
\,\, h\in \mathcal{D}(\MM)\,,\,Q\in\mathcal{P}\ .
\end{equation}
To translate it further into $T^c$ we note that
the linked cluster formula \eqref{linked-cluster} implies
\beq
T^c\big(e_\otimes^{iF}\otimes G\big)=T\big(e_\otimes^{iF}\big)^{-1}\cdot 
T\big(e_\otimes^{iF}\otimes G\big)\ ,\label{T^c-T}
\eeq
where the inverse is meant with respect to the classical product. It exists because 
$T\big(e_\otimes^{iF}\big)$ is a formal power series of the form $T\big(e_\otimes^{iF}\big)=1+\mathcal{O}(F)$.
With that we conclude that the MWI can equivalently be written in terms of $T^c$
by replacing $T$ by $T^c$ on both sides of \eqref{MWI:T}.

\paragraph{Translation of the MWI from $T^c$ into $\Gamma$.} 
Applying \eqref{Gamma} on both sides of the MWI in terms of $T^c$ we obtain
\begin{eqnarray*}
\lefteqn{\int\! dy\, h(y)\,T^c_{\mathrm{tree}}
\Big(e_\otimes^{i\Gamma(e_\otimes^S)/\hbar}\otimes \Gamma\Big(e_\otimes^S\otimes Q(y)\,\frac{\delta (S_0+S)}{\delta\varphi(y)}\Big)\Big)}\\
&\hspace{1cm}&=\int\! dy\, h(y)T^c_{\mathrm{tree}}\Big(e_\otimes^{i\Gamma(e_\otimes^S)/\hbar}\otimes \Gamma\big(e_\otimes^S\otimes
Q(y)\big)\Big)\frac{\delta S_0}{\delta \varphi(y)}\\
&\hspace{1cm}&= \int\! dy\, h(y)\,T^c_{\mathrm{tree}}\Big(e_\otimes^{i\Gamma(e_\otimes^S)/\hbar}\otimes  
\Gamma(e_\otimes^S\otimes Q(y))\frac{\delta
(S_0+\Gamma(e_\otimes^S))}{\delta \varphi(y)}\Big)\ ,
\end{eqnarray*}       
where we have used the classical MWI in terms of $T^c_\mathrm{tree}$ \eqref{MWI-tree}. It follows
\begin{equation}\label{effQAP-T}
\Gamma(e_\otimes^S\otimes Q(y))\frac{\delta(S_0+\Gamma(e_\otimes^S))}{\delta \varphi(y)}=
\Gamma\Big(e_\otimes^S\otimes Q(y)\,\frac{\delta (S_0+S)}{\delta\varphi(y)}\Big)\ .
\end{equation}
The various formulations of the MWI, in terms of $R$-products \eqref{MWI:R},  $T$-products \eqref{MWI:T}, $T^c$-products
and in terms of proper vertices \eqref{effQAP-T}, they all are equivalent.

\begin{rem}
The off-shell field equation
\beq
T\Bigl(e_\otimes^{iS/\hbar}\otimes\frac{\delta (S_0+S)}{\delta\varphi(y)}\Bigr)
=\frac{\delta S_0}{\delta\varphi(y)}\cdot T\Bigl(e_\otimes^{iS/\hbar}\Bigr)\ ,\quad\forall S\ ,
\eeq
is a further renormalization condition, which can equivalently be expressed by
\beq
\Gamma(e_\otimes^S\otimes \varphi(y))= \varphi(y)\ ,\quad\forall S\ ,
\eeq
as shown in \cite{qap}. For a $T$-product satisfying this condition and for $Q=D\varphi$
(where $D$ is a polynomial in partial derivatives) the QAP simplifies to
\begin{equation}
D\varphi(y)\,\frac{\delta(S_0+\Gamma(e_\otimes^S))}{\delta \varphi(y)}=
\Gamma\Big(e_\otimes^S\otimes D\varphi(y)\,\frac{\delta (S_0+S)}{\delta\varphi(y)}\Big)\ .
\end{equation}
\end{rem}

\subsection{The anomalous Master Ward Identity - Quantum Action Principle}

The QAP is a statement about the structure of all possible violations of Ward identities.
In our framework the main statement of the QAP is that any term violating the MWI
can be expressed as $\Gamma(\ets\otimes\Delta)$, where $\Delta$ is local (in a stronger sense
than only $\Delta\in\mathcal{F}_\mathrm{loc}$) and $\Delta=\mathcal{O}(\hbar)$
and the mass dimension of $\Delta$ is bounded in a suitable way.

\begin{thm}[Quantum Action Principle] \label{satzQAP}
(a) Let $\Gamma$ be the vertex functional belonging to a time ordered product satisfying the 
basic axioms and Translation Invariance \eqref{transinv}.
Then there exists a unique sequence of {\bf linear} maps $(\Delta^n)_{n\in\NN}$,
\beq
\Delta^n\,:\,\mathcal{P}^{\otimes (n+1)}\rightarrow\mathcal{D}^\prime(\MM,\mathcal{F}_\mathrm{loc})
\ ,\quad\otimes_{j=1}^n L_j\otimes Q\mapsto \Delta^n(\otimes_{j=1}^n L_j(x_j); Q(y))
\eeq
($\mathcal{D}^\prime(\MM,\mathcal{F}_\mathrm{loc})$ is the space of $\mathcal{F}_\mathrm{loc}$-valued
distributions on $\mathcal{D}(\MM)$), which are symmetric in the first $n$ factors,
\beq
 \Delta^n(\otimes_{j=1}^n L_{\pi j}(x_{\pi j}); Q(y))=
 \Delta^n(\otimes_{j=1}^n L_j(x_j); Q(y))\label{symmetry}
\eeq
for all permutations $\pi$, and
which are implicitly defined by the 'anomalous MWI'
\begin{equation}\label{anomMWI}
\Gamma(e_\otimes^S\otimes Q(y))\frac{\delta(S_0+\Gamma(e_\otimes^S))}{\delta \varphi(y)}=
\Gamma\Big(e_\otimes^S\otimes \Big(Q(y)\,\frac{\delta (S_0+S)}{\delta\varphi(y)}+\Delta(L;Q)(g;y)\Big)\Big)\ ,
\end{equation}
where $S=L(g)$ ($L\in\mathcal{P}\,,\,g\in\mathcal{D}(\MM)$)  and
\beq
\Delta(L;Q)(g;y):=\sum_{n=0}^\infty \frac{1}{n!}\int\! dx_1...dx_n \,
\prod_{j=1}^n g(x_j)\, \Delta^n(\otimes_{j=1}^n L(x_j); Q(y))\ .
\eeq
As a consequence of \eqref{anomMWI} the maps $\Delta^n$ have the following properties:
\begin{itemize}
\item[(i)] $\Delta^0=0$ ;
\item[(ii)] locality: there exist {\bf linear} maps $P^n_a\,:\,\mathcal{P}^{\otimes (n+1)}\rightarrow\mathcal{P}$ (where
$a$ runs through a finite subset of $(\NN_0^d)^n$), which are symmetric in the first $n$ factors,
such that $\Delta^n$ can be written as
\beq
\Delta^n(\otimes_{j=1}^n L_j(x_j);Q(y))=
\sum_{a\in (\NN_0^d)^n}\partial^a\delta(x_1-y,...,x_n-y)\,
P^n_a(\otimes_{j=1}^n L_j;Q)(y)\ .\label{Delta-local}
\eeq
\item[(iii)] $\Delta^n(\otimes_{j=1}^n L_j(x_j); Q(y))=\mathcal{O}(\hbar) \quad \forall n>0$ if $ L_j\sim \hbar^0,\ 
Q\sim\hbar^0$ .
\end{itemize}

(b) If the time ordered product satisfies the renormalization conditions Field Independence and
Scaling Degree (\ref{axiom-sd}), then each term on the r.h.s.~of \eqref{Delta-local} fulfills
\beq
|a|+{\rm dim}(P^n_a(\otimes_{j=1}^n L_j;Q))\leq \sum_{j=1}^n
{\rm dim}(L_j)+{\rm dim}(Q)+\frac{d+2}{2}-d\, n\ .\label{dim(Delta)}
\eeq
For a renormalizable interaction (that is ${\rm dim}(L)\leq d$) this
implies
\beq 
|a|+{\rm dim}(P^n_a(L^{\otimes n};Q))\leq {\rm dim}(Q)+\frac{d+2}{2}\ .
\eeq
\end{thm}
Note that \eqref{anomMWI} differs from the MWI \eqref{effQAP-T} only by the local term 
$\Delta(L;Q)(g;y)$, which clearly depends on the chosen normalization of the time ordered 
product. Therefore, $\Delta(L;Q)(g;y)=0$ is a sufficient condition for the validity of the MWI
for $Q$ and $S=L(g)$; it is also necessary due to the uniqueness of the maps $\Delta^n$.

\begin{proof}
(a) Proceeding as in Sect.~5.1, the defining relation \eqref{anomMWI} can equivalently be written in terms of $T$-products:
\beq\label{anomMWI:T} 
T\Big(e_\otimes^{iS/\hbar}\otimes  \big(Q(y)\,\frac{\delta (S_0+S)}{\delta\varphi(y)}+\Delta(L;Q)(g;y)\big)\Big)
=T\Big(e_\otimes^{iS/\hbar}\otimes Q(y)\Big)\frac{\delta S_0}{\delta \varphi(y)}\ .
\end{equation}
To $n$-th order in $g$ this equation reads
\begin{gather}
\Delta^n(L^{\otimes n};Q(y))(g^{\otimes n})=T\Big((i S/\hbar)^{\otimes n}\otimes Q(y)\Big)\cdot\frac{\delta S_0}{\delta \varphi(y)}
-T\Big((i S/\hbar)^{\otimes n}\otimes  Q(y)\,\frac{\delta S_0}{\delta \varphi(y)}\Big)\notag\\
-n\, T\Big((i S/\hbar)^{\otimes n-1}\otimes Q(y)\,\frac{\delta S}{\delta \varphi(y)}\Big)-\sum_{l=0}^{n-1} \binom{n}{l}\,
T\Big((i S/\hbar)^{\otimes n-l}\otimes \Delta^l(L^{\otimes l};Q(y))(g^{\otimes l})\Big)\ .\label{recursion-F}
\end{gather}
Taking linearity and symmetry \eqref{symmetry} into account we extend this relation to non-diagonal entries 
and write it in terms of the distributional kernels
\begin{gather}
\Delta^n(\otimes_{j=1}^n L_j(x_j);Q(y))=\Bigl(\frac{i}{\hbar}\Bigr)^n\,
T\Bigl(\otimes_{j=1}^n L_j(x_j)\otimes Q(y)\Bigr)\cdot \frac{\delta S_0}{\delta\varphi(y)}\notag\\
-\Bigl(\frac{i}{\hbar}\Bigr)^n\,T\Bigl(\otimes_{j=1}^n L_j(x_j)\otimes Q(y)
\cdot\frac{\delta S_0}{\delta\varphi(y)}\Bigr)\notag\\
-\sum_{l=1}^n \Bigl(\frac{i}{\hbar}\Bigr)^{n-1}\,T\Bigl(\otimes_{j(\not= l)}
L_j(x_j)\otimes Q(y)\sum_a(\d^a\delta)(x_l-y)\,
\frac{\d L_l}{\d(\d^a\varphi)}(x_l)\Bigr)\notag\\
-\sum_{I\subset\{1,...,n\}\,,\,I^c\not=\emptyset}\Bigl(\frac{i}{\hbar}\Bigr)^{|I^c|}\,
T\Bigl(\otimes_{i\in I^c} L_i(x_i)\otimes\Delta^{|I|}(\otimes_{j\in I} L_j(x_j); Q(y))\Bigr)
\label{recursion-x}
\end{gather}
This relation gives a unique inductive construction of the sequence $(\Delta^n)_{n\in\NN}$
(if the distribution on the r.h.s.~of \eqref{recursion-x} takes values in $\mathcal{F}_\mathrm{loc}$)
and it gives also the initial value $\Delta^0=0$. Obviously, the so obtained maps 
$\Delta^n\,:\,\mathcal{P}^{\otimes (n+1)}\rightarrow\mathcal{D}^\prime(\MM,\mathcal{F}_\mathrm{loc})$
are linear and symmetric \eqref{symmetry}. 

The main task is to prove that $\Delta^n(\otimes_{j=1}^n L_j;Q)$ (which is defined inductively by 
\eqref{recursion-x}) satisfies locality \eqref{Delta-local}; the latter implies that  
$\Delta^n(\otimes_{j=1}^n L_j;Q)$ takes values in $\mathcal{F}_\mathrm{loc}$. 
For this purpose we first prove
\beq
\supp\Delta^n(\otimes_{j=1}^n L_j;Q)\subset\DD_{n+1}\=d \{(x_1, \ldots, x_{n+1})\in 
\MM^{n+1}\,|\, x_1=\cdots =x_{n+1}\}\ ,\label{suppDelta}
\eeq
that is we show that the r.h.s.~of
\eqref{recursion-x} vanishes for $(x_1,...,x_n,y)\not\in \DD_{n+1}$. For such a configuration
there exists a $K\subset\{1,...,n\}$ with $K^c:=\{1,...,n\}\setminus K\not=\emptyset$ and either
$(\{x_k\,|\,k\in K^c\}+\bar V_+)\cap(\{x_j\,|\,j\in K\}\cup\{y\})=\emptyset$ or $(\{x_k\,|\,k\in K^c\}
+\bar V_-)\cap(\{x_j\,|\,j\in K\}\cup\{y\})=\emptyset$.
We treat the first case, the second case is completely analogous. Using causal factorization 
of the $T$-products \eqref{causfac} and locality \eqref{suppDelta} of the inductively known
$\Delta^{|I|},\,|I|<n$, we write the r.h.s.~of \eqref{recursion-x} as
\begin{gather} 
\Bigl(\frac{i}{\hbar}\Bigr)^n\Bigl(T\Bigl(\otimes_{j\in K^c} L_j(x_j)\Bigr)\star 
T\Bigl(\otimes_{i\in K} L_i(x_i)\otimes Q(y)\Bigr)\Bigr)\,\frac{\delta S_0}{\delta\varphi(y)}\notag\\ 
-T\Bigl(\otimes_{j\in K^c} L_j(x_j)\Bigr)\star\Bigl[
\Bigl(\frac{i}{\hbar}\Bigr)^n \,T\Bigl(\otimes_{i\in K} L_i(x_i)\otimes Q(y) 
\,\frac{\delta S_0}{\delta\varphi(y)}\Bigr)\notag\\ 
+\Bigl(\frac{i}{\hbar}\Bigr)^{n-1}\sum_{l\in K}T\Bigl(\otimes_{i\in K,\,i\not= l} 
L_i(x_i)\otimes Q(y)\sum_a(\d^a\delta)(x_l-y)\, 
\frac{\d L_l}{\d(\d^a\varphi)}(x_l)\Bigr)\notag\\
+\Bigl(\frac{i}{\hbar}\Bigr)^{|K^c|+|K\setminus I|}\sum_{I\subset K} 
T\Bigl(\otimes_{i\in K\setminus I} L_i(x_i)\otimes\Delta^{|I|}(\otimes_{s\in I} L_s(x_s); Q(y))\Bigr)\Bigr]\ . 
\end{gather} 
Using \eqref{hilfsid} this can be written in the form $T(\otimes_{j\in K^c} L_j(x_j))\star(...)$.
The second factor vanishes due to the validity of \eqref{recursion-x} in order $|K|$.
This proves \eqref{suppDelta}.

$\Delta^n(\otimes_{j=1}^n L_j;Q)$ is, according to its inductive definition \eqref{recursion-x},
a distribution on $\mathcal{D}(\MM^{n+1})$ which takes values in $\mathcal{F}$. Hence,
it is of the form
\begin{gather}
\Delta^n(\otimes_{j=1}^n L_j(x_j);Q(y))=\sum_k\int dz_1...dz_k\,\notag\\
f^n_k(\otimes_{j=1}^n L_j\otimes Q)
(x_1,...,x_n,y,z_1,...,z_k)\,\varphi(z_1)...\varphi(z_k)\ ,\label{Delta-F}
\end{gather}
where $f^n_k(\otimes_{j=1}^n L_j\otimes Q)(x_1,...,x_n,y,z_1,...,z_k)\in\mathcal{D}^\prime (\MM^{n+k+1})$
has the following properties:\\
- it depends linearly on $(\otimes_{j=1}^n L_j\otimes Q)$;\\
- it is invariant under permutations of the pairs $(L_1,x_1),...,(L_n,x_n)$.\\
- The distribution
$\int dx_1...dx_n dy\,f^n_k(\otimes_{j=1}^n L_j\otimes Q)
(x_1,...,x_n,y,z_1,...,z_k)\,h(x_1,...,x_n,y)$
$\in\mathcal{D}^\prime (\MM^{k})$
is symmetric under permutations of $z_1,...,z_k$ and satisfies the wave front set 
condition \eqref{WF}, for all $h\in\mathcal{D}(\MM^{n+1})$.\\
- From (\ref{recursion-x}) we see that Translation Invariance of the $T$-product \eqref{transinv} implies the same 
property for $\Delta^n$:
\beq
\beta_a\,\Delta^n(\otimes_{j=1}^n L_j(x_j);Q(y))=\Delta^n(\otimes_{j=1}^n L_j(x_j+a);Q(y+a))\ .
\eeq
Therefore, the distributions $f^n_k(\otimes_{j=1}^n L_j\otimes Q)$ depend only on the relative coordinates.

Due to \eqref{suppDelta} the support of $f^n_k(\otimes_{j=1}^n L_j\otimes Q)$ is contained in $\DD_{n+1}\times\MM^k$;
but, to obtain the assertion \eqref{Delta-local}, we have to show $\supp f^n_k(\otimes_{j=1}^n L_j\otimes Q)
\subset\DD_{n+k+1}$. For this purpose we take into account that
\beq
\frac{\delta\,T(\otimes_{j=1}^l A_j(x_j))}{\delta\varphi(z)}=0\quad\mathrm{if}\quad
z\not= x_j\,\,\,\forall j=1,...,l\ .\label{Top}
\eeq
This relation can be shown as follows: for the restriction of the time ordered product to 
$\mathcal{D}(\MM^l\setminus\DD_l)$ this property is obtained inductively by causal factorization 
\eqref{causfac}. That \eqref{Top} is maintained in the extension of the $T$-product to 
$\mathcal{D}(\MM^l)$ can be derived from 
\beq
[T(\otimes_{j=1}^l A_j(x_j))\,,\,\varphi(z)]_\star =0\quad\mathrm{if}\quad
(x_j-z)^2<0\quad\forall j=1,...,l\ ,
\eeq
which is a consequence of the causal factorization of $T(\varphi(z)\otimes
\otimes_{j=1}^l A_j(x_j))$ (cf. Sect.~3 of \cite{Epstein:1973gw}). 

Applying \eqref{Top} to the $T$-products on the
r.h.s.~of \eqref{recursion-x} and using \eqref{suppDelta}, we conclude
\beq
\supp\frac{\delta\,\Delta^n(\otimes_{j=1}^n L_j;Q)}{\delta\varphi}\subset \DD_{n+2}\ .
\eeq
It follows that the distributions $f^n_k(\otimes_{j=1}^n L_j\otimes Q)$ \eqref{Delta-F}
have support on the total diagonal $\DD_{n+k+1}$. Taking additionally Translation Invariance into account, 
we conclude that these distributions are of the form
\begin{gather}
f^n_k(\otimes_{j=1}^n L_j\otimes Q)(x_1,...,x_n,y,z_1,...,z_k)=
\sum_{a,b}C_{a,b}(\otimes_{j=1}^n L_j\otimes Q)\,\notag\\
\d^a\delta(x_1-y,...,x_n-y)\,\d^b\delta(z_1-y,...,z_k-y)\ ,\label{f-delta}
\end{gather}
where the coefficients $C_{a,b}(\otimes_{j=1}^n L_j\otimes Q)\in\CC$
depend linearly on $(\otimes_{j=1}^n L_j\otimes Q)$ and are symmetric in the first $n$ factors. 
Inserting ¸\eqref{f-delta} into \eqref{Delta-F}
we obtain \eqref{Delta-local}, the corresponding maps $P^n_a$ having the asserted properties.

The important property (iii) is obtained by taking the classical limit $\hbar\to 0$ of the 
anomalous MWI \eqref{anomMWI}: using \eqref{properinter} it results $\lim_{\hbar\to 0}
\Delta(L;Q)(g;y)=0$.

(b) The statement \eqref{dim(Delta)} is a modified version of Proposition 10(ii)
in \cite{qap}. It follows from the formulas (\cite{qap}-5.32-33) and (\cite{qap}-5.46-47) of that paper. Namely,
by using the causal Wick expansion of $\Delta^n$ (which follows from the Field Independence of the $T$-product)
and \eqref{suppDelta} it is derived in  (\cite{qap}-5.32-33) that $\Delta^n$ is of the form
\begin{gather}
\Delta^n(\otimes_{j=1}^nL_j(x_j);Q(y))=\sum_{\mathbf{l},\mathbf{a},b}C^\mathbf{l}_{\mathbf{a},b}
\,(\d^b\delta)(x_1-y,...,x_n-y)\notag\\
\cdot\prod_{i=1}^n\prod_{j_i=1}^{l_i}\Bigl(\d^{a_{ij_i}}\varphi(x_i)\Bigr)
\cdot\prod_{j=1}^l\d^{a_j}\varphi(y)\notag\\
=\sum_{\mathbf{l},\mathbf{a},b}\sum_{d\leq b}\tilde C^\mathbf{l}_{\mathbf{a},b,d}
\,(\d^d\delta)(x_1-y,...,x_n-y)\notag\\
\cdot\prod_{i=1}^n\Bigl(\d^{b_i-d_i}\prod_{j_i=1}^{l_i}\Bigl(\d^{a_{ij_i}}\varphi(y)\Bigr)\Bigr)
\cdot\prod_{j=1}^l\d^{a_j}\varphi(y)\ ,
\end{gather}
where $\mathbf{l}\equiv(l_1,...,l_n;l),\,\mathbf{a}\equiv (a_{11},...,a_{1l_1},...,a_{n1},...,a_{nl_n}
;a_1...a_l)$ and $C^\mathbf{l}_{\mathbf{a},b}\,,\,\tilde C^\mathbf{l}_{\mathbf{a},b,d}$
are numerical coefficients which depend also on $(L_1,...,L_n,Q)$. Since the $T$-product satisfies the 
axiom Scaling Degree the range of $b$ is bounded by (\cite{qap}-5.46). The l.h.s.~of
\eqref{dim(Delta)} is given by
\beq
|d|+|b-d|+\sum_{i=1}^n\sum_{j_i=1}^{l_i}\Bigl(|a_{ij_i}|+\frac{d-2}{2}\Bigr)
+\sum_{j=1}^{l}\Bigl(|a_j|+\frac{d-2}{2}\Bigr)\ ,
\eeq
which agrees with the l.h.s.~of (\cite{qap}-5.47). Hence, it is bounded by the r.h.s.~of (\cite{qap}-5.47).
\end{proof}

\begin{rem}
Since the $T$-product $T(F^{\otimes n})$ depends only on the (local) functional $F$
and not on how $F$ is written as $F=\sum_k\int dx\,g_k(x)\,P_k(x)\,\,(g_k\in\mathcal{D}(\MM),\ 
P_k\in\mathcal{P})$, we conclude from \eqref{recursion-x} that we may express the violating term 
$\Delta(L;Q)(g;y)$  as follows: given $A=\int dx\, h(x)\,Q(x)\,\delta S_0/\delta\varphi(x)$
($h\in\mathcal{D}(\MM),\ Q\in\mathcal{P}$), there exists a linear and symmetric map
$\Delta_A\,:\,\TT\mathcal{F}_\mathrm{loc}\rightarrow \mathcal{F}_\mathrm{loc}$ which is 
uniquely determined by
\beq
\Delta_A(e_\otimes^{L(g)})\=d \int dy\, h(y)\,\Delta(L;Q)(g;y)\ .
\eeq
A glance at \eqref{recursion-x} shows that $\Delta_A$ depends linearly on $A$. The corresponding smeared 
out version of the QAP is given in \cite{qap}.
\end{rem}

We are now going to reformulate our version of the QAP (Theorem \ref{satzQAP}) in the form given in the 
literature. Motivated by (\ref{properinter}), we interpret $\Gamma_\mathrm{tot}(S_0,S)
\defi S_0+\Gamma(e_\otimes^S)$ as the \emph{proper total action} associated with the  
classical action $S_{\mathrm{tot}}=S_0+S$. For $P\in \mathcal{C}^\infty (\MM,\mathcal{P})$ the 
'insertion' of $P(x)$ into $\Gamma_\mathrm{tot}(S_0,S)$ is denoted and defined by\footnote{The 
dot does not mean the classical product here!}
\beq  
P(x)\cdot \Gamma_\mathrm{tot}(S_0, S)\defi 
\frac{\delta}{\delta\rho(x)}\Big|_{\rho\equiv 0}\Gamma_\mathrm{tot}\Big(S_0, S+\int\! 
dx \rho(x) P(x)\Big) = \Gamma\big(e_\otimes^S\otimes P(x)\big)\ ,  \label{insertion}
\eeq  
where $\rho\in \mathcal{D}(\MM)$ is an 'external field'. Setting
$S'\defi S+\int\! dx \rho(x) Q(x)$ and introducing the local field
\beq
\Delta(x)\defi Q(x)\,\frac{\delta(S_0+S)}{\delta \varphi(x)}+\Delta(L;Q)(g;x)
\in\mathcal{C}^\infty (\MM,\mathcal{P})\ ,\label{qap-insertion}
\eeq
the anomalous MWI \eqref{anomMWI} can be rewritten as 
\beq\label{QAP} 
\frac{\delta \Gamma_\mathrm{tot}(S_0, S')}{\delta \rho(x)}
\frac{\delta \Gamma_\mathrm{tot}(S_0, S')}{\delta \varphi(x)}\Big|_{\rho\equiv 0}
=\Delta(x)\cdot \Gamma_\mathrm{tot}(S_0, S)\ .
\eeq 
The $\hbar$-expansion of the right-hand side starts with 
\beq 
\Delta(x)\cdot \Gamma_\mathrm{tot}(S_0, S)=Q(x)\frac{\delta(S_0+S)}{\delta \varphi(x)}+\mathcal{O}(\hbar)
\equiv \frac{\delta (S_0+S')}{\delta \rho(x)}\frac{\delta (S_0+S')}{\
\delta \varphi(x)}\Big|_{\rho= 0}+\mathcal{O}(\hbar),\label{QAP-hbar} 
\eeq 
where \eqref{properinter} is used. To discuss the mass dimension of the local insertion $\Delta$ \eqref{qap-insertion},
we assume that there is an open region $\emptyset\not=\mathcal{U}\subset\MM$ such that the test function $g$ which switches 
the interaction is constant in $\mathcal{U}$:  $g\vert_\mathcal{U}=\mathrm{constant}$.
For $x\in\mathcal{U}$ the insertion $\Delta(x)$ is a field polynomial with constant coefficients. By ${\rm dim}(\Delta)$ 
we mean the mass dimension of this polynomial. For a renormalizable interaction Theorem \ref{satzQAP}(b) implies
\beq
{\rm dim}(\Delta)\leq\mathrm{dim}(Q)+\frac{d+2}{2}=\mathrm{dim}(Q)-\mathrm{dim}(\varphi)+d\ . 
\label{QAP-dim}
\eeq
This version (\ref{QAP})-(\ref{QAP-dim}) of the QAP, which we have proved in the framework of CPT,
formally agrees with the literature, namely with the 'QAP for nonlinear variations of the fields' 
(formulas (3.82)-(3.83) in \cite{Piguet:1995er}). This is the most important and most 
difficult case of the QAP.

As explained in \eqref{klfeldgloff}, the MWI reduces for $Q=1$ to the off-shell field equation.
Setting $Q=1$ in (\ref{QAP})-(\ref{QAP-dim}) and using $\Gamma(\ets\otimes 1)=1$, we obtain
$\delta \Gamma_\mathrm{tot}(S_0, S)/\delta \varphi(x)=\Delta(x)\cdot \Gamma_\mathrm{tot}(S_0, S)\,$, 
where $\Delta(x)\cdot \Gamma_\mathrm{tot}(S_0, S)=\delta (S_0+S)/\delta \varphi(x)++\mathcal{O}(\hbar)$
and ${\rm dim}(\Delta)\leq d-\mathrm{dim}(\varphi)$, which formally agrees with formulas (3.80)-(3.81) 
in \cite{Piguet:1995er}. The latter are called there the 'QAP for the equations of motion', as 
expected from \eqref{klfeldgloff}.

\begin{rem}
An 'insertion' \eqref{insertion} being a rather technical notion, the violating term 
$\Gamma\Big(e_\otimes^S\otimes \Delta(L;Q)(g;y)\Big)$ in the anomalous MWI \eqref{anomMWI}
can be much better interpreted by writing  \eqref{anomMWI} in terms of $R$-products:
\begin{equation} 
R\Bigl(e_\otimes^S\otimes Q(y)\frac{\delta(S_0+S)}{\delta \varphi(y)}\Bigr)+
R(e_\otimes^S\otimes\Delta(L;Q)(g;y))= 
R(e_\otimes^S\otimes Q(y))\,\frac{\delta S_0}{\delta\varphi(y)}\ . \label{violatingterm}
\end{equation} 
In this form, the violating term $R(e_\otimes^S\otimes\Delta(L;Q)(g;y))$ is the {\it interacting field}
to the interaction $S$ and belonging to the local field $\Delta(L;Q)(g;y)$.
\end{rem}

\section{Algebraic renormalization}

In this section we sketch, for the {\it non-expert reader}, the crucial role of the QAP in algebraic 
renormalization. For shortness, we strongly simplify.

In algebraic renormalization one investigates, whether violations of Ward identities can be removed 
by finite renormalizations of the $T$-products. The results about the structure of the violating term given by the 
QAP are used as follows.
\begin{itemize}
\item Algebraic renormalization starts with the anomalous MWI \eqref{anomMWI}, that is the result that the 
MWI can be violated only by an insertion term, i.e.~a term of the form $\Gamma\Big(e_\otimes^S\otimes \Delta\Big)$
for some $\Delta\in\mathcal{F}_\mathrm{loc}$, cf.~\eqref{violatingterm}.
\item Algebraic renormalization proceeds by induction on the order of $\hbar$. 
To start the induction one uses
that $\Delta\equiv\Delta(L;Q)(g;y)$ is of order $\mathcal{O}(\hbar)$.
\item Because the finite renormalization terms, which one may add to a $T$-product, must be
{\it local} (in the strong sense of \eqref{Delta-local})
and compatible with the axiom Scaling Degree, it is of crucial importance that $\Delta(L;Q)(g;y)$ 
satisfies locality \eqref{Delta-local} and the bound \eqref{dim(Delta)} on its mass dimension.
\end{itemize}
For many Ward identities it is possible to derive a {\it consistency equation} for $\Delta(L;Q)(g;y)$.
Frequently this equation can be interpreted as the statement that $\Delta(L;Q)(g;y)$ is a {\it cocycle} in the cohomology
generated by the corresponding symmetry transformation $\delta$ acting on some space $\mathcal{K}\subset
\mathcal{F}_\mathrm{loc}$. For example, $\delta$ is a nilpotent derivation (as the BRST-transformation\footnote{The 
cohomological structure of BRST-symmetry is much richer as mentioned here, see \cite{Henneaux:1992ig}.})
or a family of derivations $(\delta_a)_{a=1,...,N}$ fulfilling a Lie algebra relation $[\delta_a,\delta_b]=
f_{abc}\,\delta_c$.

If the cocycle $\Delta(L;Q)(g;y)$ is a {\it coboundary}, it is usually possible to remove this violating term
by a finite renormalization. Hence, in this case, 
the solvability of the considered Ward identity amounts to the question whether this cohomology is trivial.
For a renormalizable interaction the bound \eqref{dim(Delta)} on the mass dimension makes it possible to reduce 
the space $\mathcal{K}$ to a finite dimensional space, this simplifies the cohomological question enormously.

Many examples for this pattern are given in \cite{Piguet:1995er}. In the framework of CPT the QAP
and its application in algebraic renormalization have been used 
to prove the Ward identities of the $O(N)$ scalar field model \cite{qap} (as a simple example to illustrate how 
algebraic renormalization works in CPT) and, much more relevant, BRST-symmetry of Yang-Mills fields in 
curved space-times \cite{Hollands:2007}.


\subsection*{Acknowledgment}
We very much profitted from discussions with Klaus Fredenhagen. We are grateful also to 
Raymond Stora for valuable and detailed comments. While writing this paper, M.D.~was
a guest of the Institut for Theoretical Physics of the University of G\"ottingen, he 
thanks for hospitality.
\end{document}